# Long Living Hot and Dense Plasma from Relativistic Laser-Nanowire Array Interaction

Ehsan Eftekhari-Zadeh[1,2,3,a)], Mikhail Gyrdymov[4], Parysatis Tavana[1,4,5], Robert Loetzsch[1,2], Ingo Uschmann[1,2], Thomas Siefke[6,7,8], Thomas Käsebier[6], Uwe Zeitner[6,7], Adriana Szeghalmi[6,7], Alexander Pukhov[9], Dmitri Serebryakov[10], Evgeni Nerush[10], Igor Kostyukov[10], Olga Rosmej[4,5], Christian Spielmann[1,2,3], Daniil Kartashov[1,2,3,b)]

[1] Institute of Optics and Quantum Electronics, Friedrich Schiller University Jena, 07743 Jena, Germany
[2] Helmholtz Institute Jena, 07743 Jena, Germany
[3] Abbe Center of Photonics, Friedrich Schiller University Jena, 07745 Jena, Germany
[4] GSI Helmholtzzentrum für Schwerionenforschung GmbH, 64291 , Germany
[5] Goethe-Universität Frankfurt am Main, 60438 Frankfurt am Main, Germany
[6] Institute of Applied Physics, Friedrich Schiller University Jena, 07745 Jena, Germany
[7] Fraunhofer Institute for Applied Optics and Precision Engineering IOF, 07745 Jena, Germany
[8] University of Applied Sciences, Carl-Zeiss-Promenade 2, 07745 Jena, Germany
[9] Institut für Theoretische Physik, Heinrich-Heine-Universität Düsseldorf, 40225 Düsseldorf, Germany
[10] Institute of Applied Physics RAS, 603950 Nizhny Novgorod, Russia

*Correspondence should be addressed to: a) e.eftekharizadeh@uni-jena.de, b) daniil.kartashov@uni-jena.de*

## Abstract:

Long-living, hot and dense plasmas generated by ultra-intense laser beams are of critical importance for laser-driven nuclear physics, bright hard X-ray sources, and laboratory astrophysics. We report the experimental observation of plasmas with nanosecond-scale lifetimes, near-solid density, and keV-level temperatures, produced by irradiating periodic arrays of composite nanowires with ultra-high contrast, relativistically intense femtosecond laser pulses. Jet-like plasma structures extending up to 1 mm from the nanowire surface were observed, emitting K-shell radiation from He-like Ti$^{20+}$ ions. High-resolution X-ray spectra were analyzed using 3D Particle-in-Cell (PIC) simulations of the laser–plasma interaction combined with collisional–radiative modeling (FLYCHK). The results indicate that the jets consist of plasma with densities of $10^{20}$–$10^{22}$ cm$^{-3}$ and keV-scale temperatures, persisting for several nanoseconds. We attribute the formation of these jets to the generation of kiloTesla-scale global magnetic fields during the laser interaction, as predicted by PIC simulations. These fields may drive long-timescale current instabilities that sustain magnetic fields of several hundred tesla, sufficient to confine hot, dense plasma over nanosecond durations.

# 1- Introduction

Structuring solid target surfaces on the micro- and nanoscale significantly enhances laser energy coupling into the target bulk, leading to volumetric heating and the formation of high-energy-density plasmas over relatively large volumes [1]. In contrast, conventional flat targets confine dense, hot plasmas to a sheet with thickness on the order of the skin depth (typically tens of nanometers), while a considerable fraction of the laser energy is absorbed in the near-critical preplasma. Surface nanostructuring enables deep penetration and near-complete absorption of laser energy within the target, producing solid-density plasmas that extend to micron-scale depths. This results in remarkable enhancement of particle fluxes (electrons and ions) as well as X-ray yields [1–17]. Among various nanostructured morphologies, aligned nanowire arrays (NWAs) have attracted particular attention over the past decade. They exhibit exceptionally high absorption efficiency, support volumetric heating, and facilitate the generation of ultra-hot and ultra-dense plasmas [1, 7, 13–18]. Numerous studies have demonstrated that NWAs outperform flat targets in generating emissions of X-rays [8, 19, 20], gamma rays [21], electrons [12, 18, 22], ions [10, 23, 24], and neutrons [25].

A distinctive feature of NWA morphology is the ability of relativistically intense laser pulses to drive highly directed currents along individual nanowires. Electrons accelerated to ultra-relativistic energies by the laser ponderomotive force in the skin layer form forward currents, which in turn induce compensating return currents within the wire bulk to maintain charge neutrality. These return currents can reach densities of ~100 MA/µm², with total currents exceeding the Alfvén vacuum limit by an order of magnitude, generating magnetic fields of tenths of kiloTesla strength [12, 26]. The resulting Z-pinch effect may further compress the plasma, increasing both density and temperature by an additional order of magnitude [26].

Despite several experimental studies, relativistic laser–NWA interactions remain a central topic in high-energy-density physics due to their strong potential for advancing fundamental science and enabling novel applications. However, comprehensive investigations remain limited, owing to stringent experimental requirements. These include the need for ultra-high temporal contrast ($<10^{-10}$ at picosecond timescales) in high-power laser systems and the technological challenges of fabricating large-area, well-ordered NWAs with controlled parameters (diameter, spacing, length) and extreme aspect ratios ($\gtrsim 10^3$). In particular, systematic studies combining complex diagnostics of both particles and radiation in a single experimental campaign are still lacking.

Here, we present a combined experimental and numerical investigation of relativistic interactions between intense, ultra-high contrast femtosecond laser pulses and composite nanowire array targets consisting of a low-Z Si core (Z = 14) and a mid-Z Ti shell (Z = 22). Using a suite of diagnostics, we simultaneously measured X-ray and particle spectra and performed X-ray source imaging. For comparison, we also studied planar reference targets of similar composition. High-resolution crystal spectrometers enabled precise measurements of characteristic Ti X-ray emission lines, allowing us to estimate key plasma parameters such as bulk electron temperature and density. An imaging crystal spectrometer provided spatially resolved measurements of Ti X-ray emission along the laser propagation axis. A particularly intriguing finding was the observation of ≈1 mm-long jet-like plasma structures, emitting the He$_\alpha$ line of Ti$^{20+}$ ions, extending from the target surface toward the laser. These structures were observed exclusively with NWA targets. Finally, we measured and compared the energy distributions of electrons and ions emitted from both the front and rear sides of NWA and planar targets.

The paper is organized as follows: Sec. II describes the experimental setup, materials, and methods. Sec. III presents the results and discussion, including comparisons with FLYCHK and 3D PIC simulations. Sec. IV concludes with a summary of the key findings.

## 2- Experimental Setup – Materials and Methods

The experiments were carried out at the multi-terawatt JETI 40 laser system (University of Jena), delivering up to 0.7 J in 30 fs pulses at a central wavelength of 0.8 µm. The general layout of the setup is shown in Fig. 1(a). The temporal contrast of the pulses, measured with a third-order autocorrelator (Sequoia), was in the range $10^{-10}$–$10^{-6}$ for delays from nanoseconds down to 50 ps [Fig. 1(b)]. To further improve the contrast, the output beam was frequency doubled in a 700-µm-thick, type-I KDP crystal, achieving ≈25% conversion efficiency and yielding ~150 mJ energy at 0.4 µm. The second-harmonic pulse duration was estimated to be ≈ 40 fs using SNLO simulations [27]. Frequency upconversion to the second harmonic provided an additional contrast enhancement of ≈ $10^6$ [28], thereby ensuring that the nanostructures interacted predominantly with the main pulse. Residual 800-nm radiation was removed using three broadband dichroic mirrors, highly reflective (> 99%) at 400 nm and highly transmissive (> 99%) at 800 nm.

A 5 µm-thick pellicle beam splitter with 5% reflectivity was used to direct a small fraction of the beam to a power meter located outside the chamber for single-shot energy monitoring [Fig. 1(a)]. Except for the CCD detector for the GaAs spectrometer, all diagnostics were placed inside a vacuum chamber maintained at $2 \times 10^{-4}$ mbar. The second-harmonic beam, 60 mm in diameter, was focused under normal incidence onto the targets using a 3-inch, f/2.3, aluminum-coated off-axis parabolic mirror with an effective focal length of 177.5 mm. To optimize the focal spot, an adaptive mirror controlled by a genetic algorithm was positioned upstream of the compressor in the beamline. The measured focal spot had an average full-width at half-maximum (FWHM) of 2.6 µm [Fig. 1(c)]. The focal-plane intensity distribution was characterized using a CCD camera with a 20× microscope objective and subsequently numerically integrated. Assuming a Gaussian temporal profile of 40 fs FWHM, the peak intensity was estimated to exceed $3 \times 10^{19}$ W/cm², corresponding to a relativistic parameter of $a_0 \approx 2$.

The NWA targets are periodic arrays of composite L ≈ 5 µm long nanowires consisting of a 100 nm diameter Si core and a 25 nm thick TiO$_2$ cladding, resulting in the total diameter of D = 150 nm (see Fig. 2 and supplementary materials for manufacturing details).

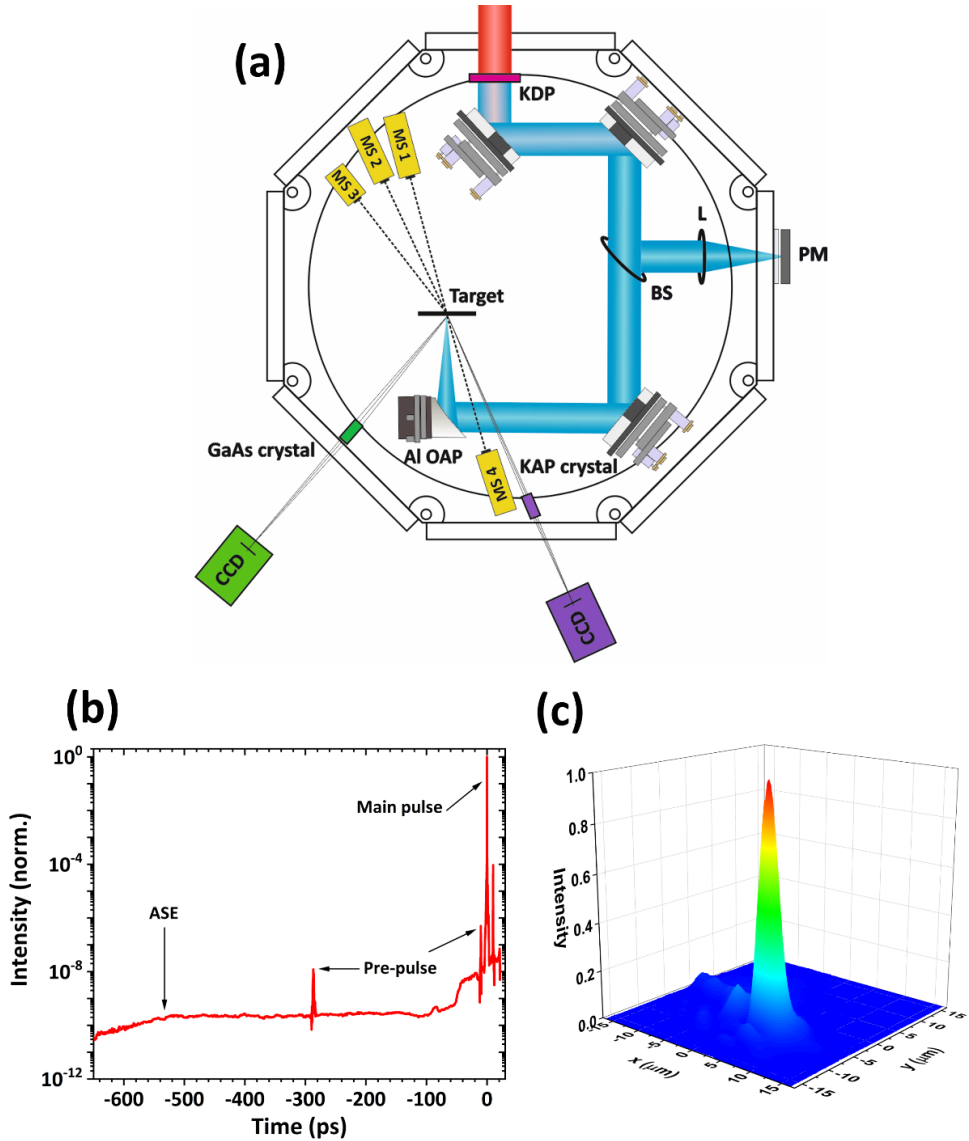

**Fig. 1.** a) The sketch of the experimental setup. KDP – the frequency doubling KDP crystal, BS – the pellicle 5% beam splitter, L – the lens focusing the reflected light into the power meter head PM for single shot energy measurements, OAP – off-axis parabolic mirror, MS1, MS2 and MS4 are ion and electron spectrometers, MS3 is the electron spectrometer, GaAs and KAP crystals are used to measure the X-ray emission spectrum of Ti and Si correspondingly with CCD cameras as detectors. b) Measured temporal contrast in the laser pulse before the frequency doubling. c) Measured intensity distribution in the focal spot detected with the CCD camera.

The period is S = 400 nm (center-to-center); see Fig. 2.. As reference targets, we employed the same thickness Si substrate coated by the 25 nm $TiO_2$ film (hereafter referred to as the reference target) and a 25 µm thick Ti foil. It is noteworthy that the Rayleigh length of ≈ 12 µm for the focused laser beam is longer than the length of the nanowires. The arrays were etched into individual 8 mm diameter, 50 µm thick Si membranes allowing multiple shots per sample and yielding reproducible X-ray and particle spectra. The experiments were conducted in single-shot mode, and the particle and X-ray spectra were accumulated over 10 shots.

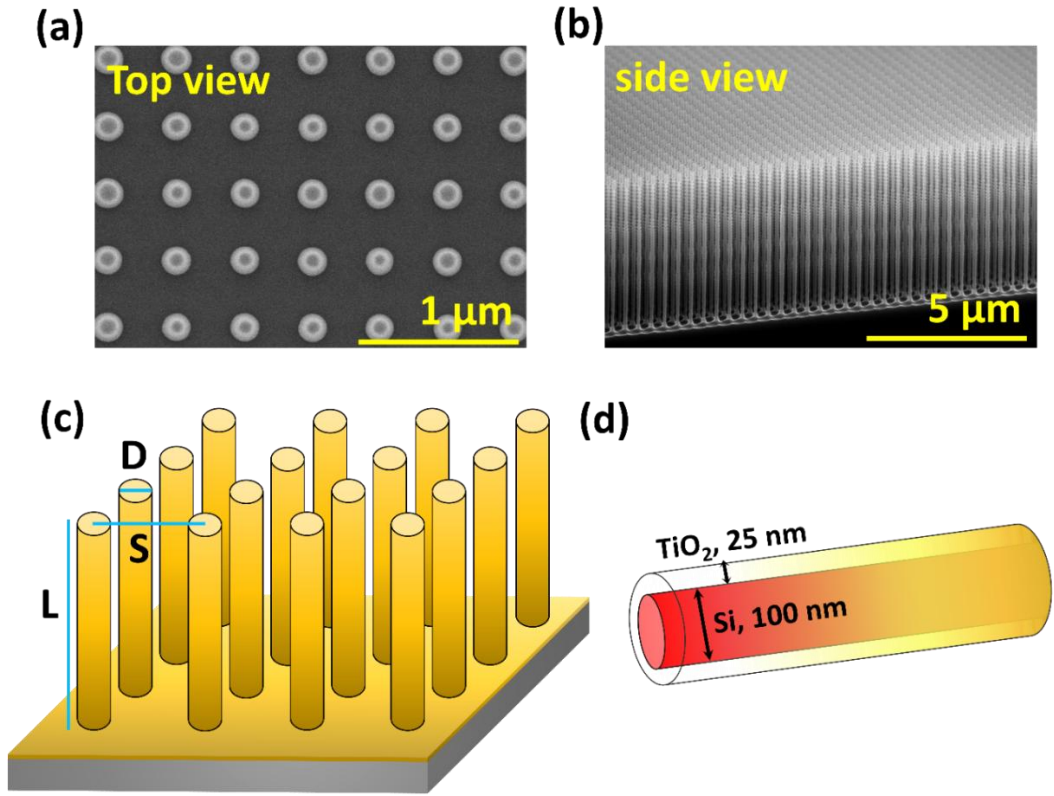

**Fig. 2.** SEM images of top (a) and side (b) views of a vertically aligned nanowire array target. c) Schematic diagram of nanowire array: L, D, and S are the nanowire length, diameter, and spacing, respectively. d) Schematic of a single nanowire illustrating the material composition of nanowires.

A comprehensive set of diagnostics was employed to simultaneously record X-ray, electron, and proton spectra. For X-ray measurements, two crystal spectrometers equipped with CCD detectors were used to measure K-shell line emissions ranging from $K_\alpha$ to $K_\beta$, extending up to Lyman lines for Si and Ti. A flat KAP crystal spectrometer coupled to a CCD covered the range 1.7–2.5 keV, corresponding to Si K-line emission from $Si^+$ to $Si^{13+}$ [Fig. 1(a)]. For Ti emission, an imaging crystal spectrometer based on a toroidally bent GaAs (111) crystal with horizontal and vertical radii of 1600 mm and 101 mm, respectively, was employed [29, 30]. In combination with CCD detection, this setup covered the range 4.5–4.95 keV, corresponding to Ti K-line emission from $Ti^+$ to $Ti^{21+}$, with a spectral resolution of $\approx 3500$ ($\approx 2$ eV). The CCD camera for this diagnostic was positioned outside the chamber, separated by a 50 µm Kapton window. In addition to spectral information, this spectrometer provided spatial imaging of the X-ray source with $\approx 5$ µm resolution.

Particle spectra were measured with permanent-magnet spectrometers equipped with image plates (IPs) as detectors [31]. Two spectrometers with a magnetic flux density of 990 mT were positioned 25 cm behind the target at angles of 20° and 30° relative to the target normal (MS1 and MS2 in Fig. 1(a)). These instruments measured electron spectra from 1.4–100 MeV and proton spectra from 0.1–10 MeV, respectively. Electrons were recorded using BASF MS IPs, whose response to incident electrons as a function of energy and incidence angle was taken from [32, 33]. Protons were detected on BASF TR IPs shielded by a 15 µm Al foil to block all ions except protons. In addition, a third spectrometer (MS3) with a magnetic flux density of 250 mT was placed 25 cm from the target at a 45° angle relative to the normal, measuring electron spectra in the range 0.1–10 MeV. On the front side of the target, another spectrometer of the same type as MS1 and MS2 was located 22 cm away at 11° relative to the normal, allowing simultaneous measurements of electron and proton spectra [Fig. 1(a)].

## 3- Experimental Results

### 3.1 Particle Spectra

The measured electron spectra at the front and rear sides of the targets are presented in Fig. 3. Overall, both sides of the nanowire targets yield 4–5 times more hot electrons compared to the reference target and the foil. In addition, the hot-electron temperature in the 1.6–2.5 MeV energy range is up to 2.5 times higher for the nanowires. At the rear side of the reference and NWA targets, the angular distribution of electrons in the 1.6–2.5 MeV range is relatively uniform, with comparable particle counts at detection angles of 20°, 30°, and 45° (Fig. 3a–c). In contrast, the foil target shows no detectable electrons at 45° (not shown). We attribute this difference to the target thickness (50 µm Si substrate for the reference and NWA targets versus 25 µm for the Ti foil) combined with the relatively low hot-electron temperature, estimated to be 0.57 MeV using the Wilks scaling for the laser parameters employed in our experiments.

The proton energy spectra measured at a 20° detection angle from the rear side of the targets are shown in Fig. 4a. For the Ti foil, the spectrum exhibits a cut-off around 0.4 MeV, which is reasonably close to the ~1 MeV estimate obtained using the TNSA acceleration model and the formulas in [34]. No proton signal was detected from the reference target. This absence can be attributed to its twice the thickness compared to the foil, which reduces the accelerating electric field strength by approximately a factor of four [34], thereby lowering the proton energies below the spectrometer detection threshold. In contrast, the NWA target—despite having the same Si substrate thickness as the reference target—produces even higher-energy protons than the foil. This enhancement can be explained by the higher temperature and density of electrons generated in the NWA, which compensates for the $d^{-2}$ scaling of the accelerating field with substrate thickness [34].

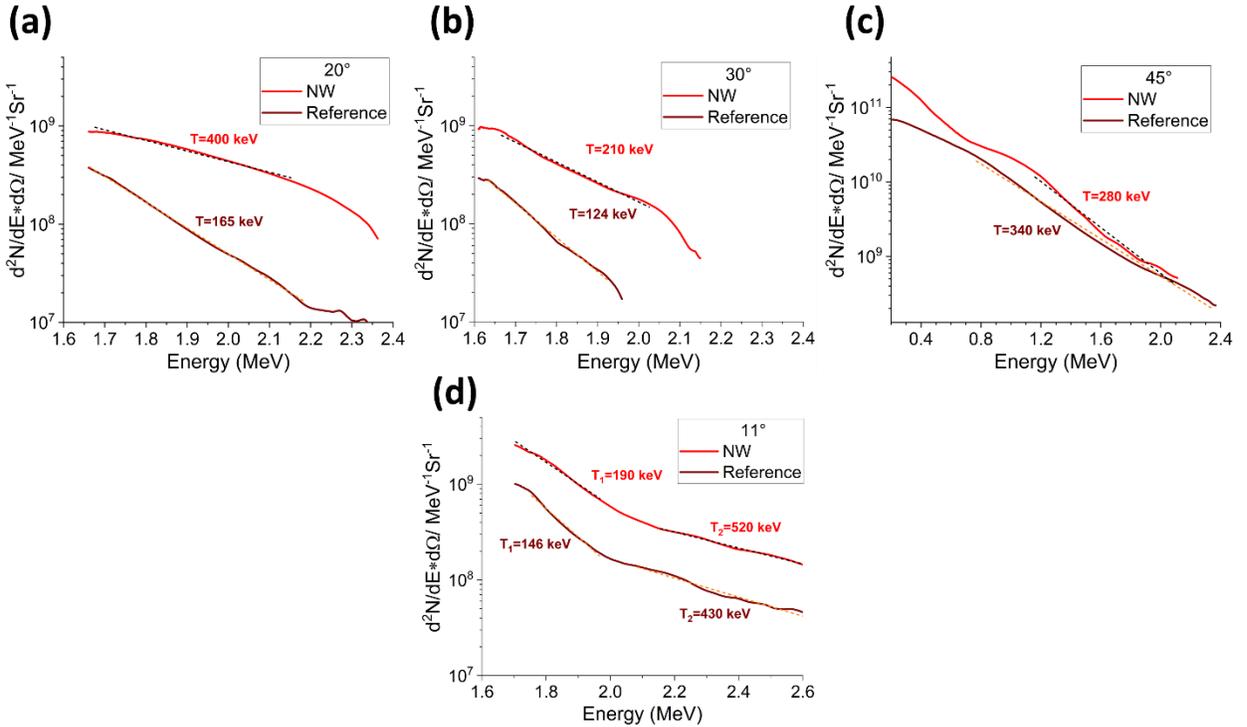

**Fig. 3.** Electron energy spectra measured at the rear side of the targets (a–c) and at the front side (d). Results are shown for nanowire array targets (red curves) and for the reference planar target (wine-colored curves).

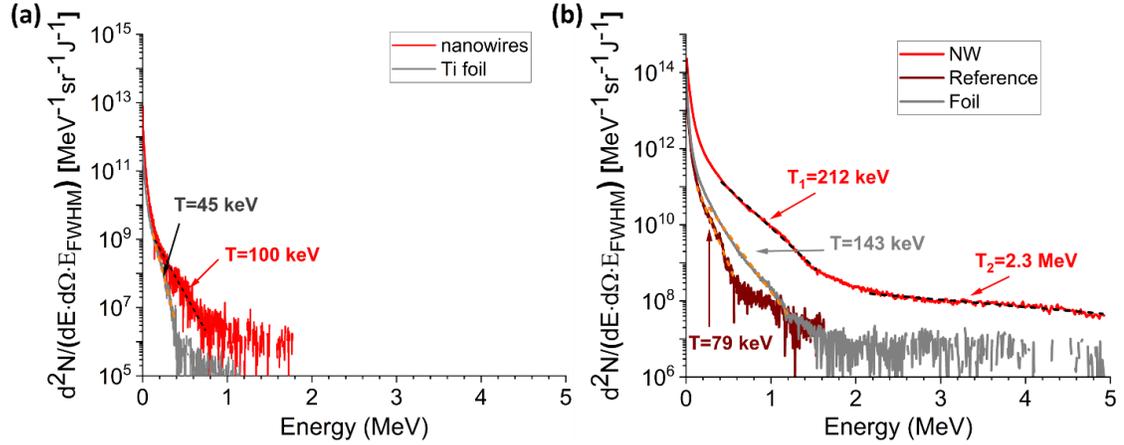

**Fig. 4.** Measured proton spectra from the (a) rear side and (b) front side of nanowire array targets (red line), reference target (wine line) and the Ti foil (gray line).

A more pronounced difference is observed in the proton spectra measured at the front side of the targets (Fig. 4b). The foil and reference targets exhibit cut-off energies of 1.2 MeV and 0.6 MeV, respectively. However, the NWA targets show an enhancement of more than two orders of magnitude in the proton yield above 1 MeV, with cut-off energies exceeding the spectrometer limit of 5 MeV. This highly efficient front-side ion acceleration is consistent with the TNSA mechanism at NWA surfaces recently reported in [25]. Notably, Ref. [25] demonstrated backward proton acceleration up to 7.5 MeV at the NWA front surface using laser pulses of intensity $3\times10^{21}$ W/cm$^2$ at 0.4 µm wavelength, whereas in our experiments similar proton energies (> 5 MeV) are obtained with laser pulses two orders of magnitude less intense. This suggests that, in contrast to rear-side TNSA—where the accelerating field scales nearly linearly with laser intensity and is governed by hot electrons directly accelerated by the laser [34]—the backward-directed front-side TNSA, driven by the return current, exhibits strong saturation of the acceleration efficiency with increasing laser intensity.

### 3.2 X-ray spectroscopy

The measured K-shell X-ray emission spectra from NWA targets show a pronounced intensity enhancement compared to reference planar targets (Fig. 5a, b). In the Si K-shell spectral range (1.7 – 2.5 keV), both $K_\alpha$ and $K_\beta$ lines are observed up to the H-like state (Ly$_\alpha$) of Si$^{13+}$ (Fig. 5a). Emission from the He-like Si$^{12+}$ and H-like Si$^{13+}$ charge states is more than six times stronger for NWA targets than for reference planar targets.

Similarly, Ti K-shell emission in the 4.5–4.95 keV range includes $K_\alpha$ and $K_\beta$ lines up to the H-like ($Ly_\alpha$) state of $Ti^{21+}$ (Fig. 5b). Emission from low Ti charge states up to B-like $Ti^{17+}$ can be attributed to optical field ionization (OFI) driven directly by the laser field, as estimated using ADK theory [35], whereas ionization to higher charge states up to $Ti^{21+}$ results from electron-impact ionization in collisions with plasma electrons [36]. Notably, Ti is the heaviest element for which temperature-dependent dielectronic satellite (DS) lines can be spectrally resolved from the He-like intercombination line y [36]. In the measured spectra, the He-like $Ti^{20+}$ lines include the resonance $He_\alpha(w)$ transition ($1s^2 - 1s2p$, 4.74 keV), the intercombination $He_\alpha(y)$ transition ($1s^2 - 1s2p\ ^3P_1$, 4.72 keV), and associated DS transitions ($1s^2 2l - 1s2l'2l$ with l, l' = s, p). The emission yield from He-like $Ti^{20+}$ shows an enhancement exceeding two orders of magnitude when using NWA targets compared to either the planar reference target or the conventional foil (Fig. 5b).

The appearance of X-ray emission from high charge states (He- and H-like ions) is a clear indicator of hot, dense plasma [1,36]. The exclusive observation of H-like $Ti^{21+}$ $Ly_\alpha$ emission in NWA targets, together with the enhanced He-like $Ti^{20+}$ yield, demonstrates the generation of a volumetric plasma with higher temperature and density than in planar targets. This volumetric plasma remains hot (several keV) for up to ≈ 1 ps after the laser peak, during which it expands and cools both hydrodynamically and radiatively, thereby producing enhanced X-ray emission [8].

The spectra in Fig. 5 are spatially integrated over the CCD image. The arrangement of the toroidal crystal for Ti emission line measurements at ≈ 45° relative to the target normal (see Fig. 1a) enables one-dimensional (1D) imaging of the emission source in the backward direction (see supplementary materials for details of image construction). Spatially resolved X-ray line emission spectra from the reference and NWA targets are shown in Fig. 6. The horizontal axis corresponds to the diffraction (spectral) direction, while the vertical axis provides a 1D image of the emission source at the corresponding photon energy with ≈4× magnification and ≈10 μm spatial resolution.

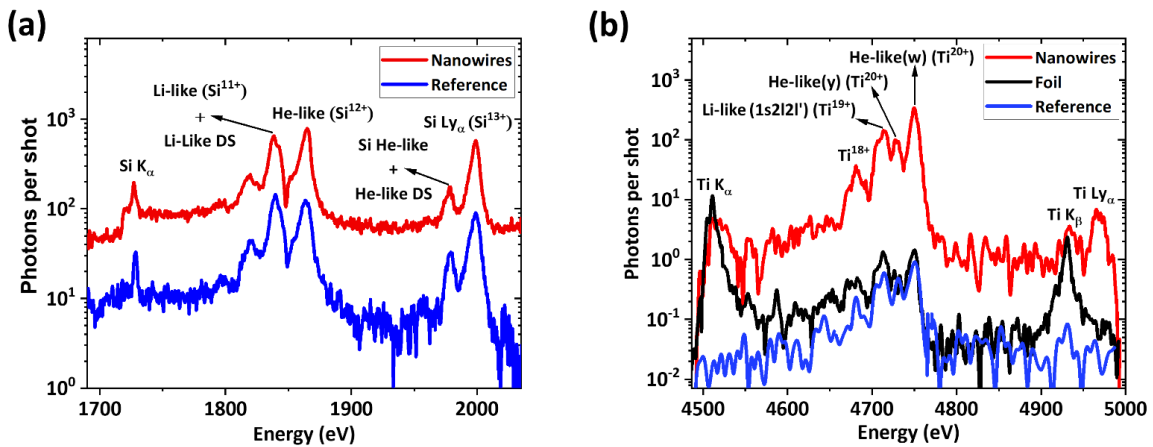

**Fig. 5.** The measured X-ray emission spectra from the nanowire array targets (solid red lines) and reference planar targets (solid blue lines) for (a) Si and (b) Ti ions. The black solid line in (b) is the measured X-ray lines from Ti foil used as a reference for $K_\alpha$ and $K_\beta$ lines for calibration.

For the reference target, only emission lines from Li-like ($Ti^{19+}$) and He-like ($Ti^{20+}$) ions are observed, localized at the surface within a ≈20 µm FWHM spot—comparable to the resolution limit of the imaging system (Fig. 6a). In contrast, NWA targets produce jet-like plasma structures extending up to 1 mm from the surface, emitting in transitions of He-like $Ti^{20+}$ ions (Fig. 6b). The spatial distributions of the "cold" Kα line emission and of the He-like and H-like emission from $Ti^{20+}$ and $Ti^{21+}$ ions in NWA targets are shown in Fig. 6c. These distributions were obtained by spectrally integrating each line as a function of the vertical coordinate in Fig. 6b.

Analysis of the 1D imaging with the toroidal crystal at 45° to the surface normal indicates that the symmetric ≈ 150 µm (foot-to-foot) extent of the "cold" Kα emission represents the source size at the surface (Fig. 6c, top). The He-like emission shows a symmetric ≈ 100 µm component corresponding to the surface source, together with an asymmetric "jet" extending up to ≈ 0.9 mm from the surface (Fig. 6c, middle). The H-like emission originates from a ≈ 50 µm surface spot and a ≈ 120 µm jet. The $Ti^{20+}$ $He_\alpha$ (w) line, corresponding to the $1s^2 \rightarrow 1s2p$ transition, has an energy of 4.75 keV and a radiative decay rate of $2.4 \times 10^{14}$ $s^{-1}$, corresponding to a ≈ 4 fs radiative lifetime [37]. Thus, continuous electron collisional pumping is required for this emission to persist on longer timescales. This, in turn, implies that a keV-temperature, high-density plasma is sustained over timescales sufficient to produce a millimeter-scale jet.

To estimate the lifetime of such plasma, we consider two possible scenarios. In the first, as discussed above, our measurements of the proton spectra indicate highly efficient backward ion acceleration via the TNSA mechanism at the nanowire tips. As an extreme case, one might assume that $Ti^{20+}$ ions are accelerated together with protons. For protons with energies above 5 MeV, the velocity exceeds $3 \times 10^9$ cm/s (≈ 0.1 c). Given the charge-to-mass ratio of 0.42 for $Ti^{20+}$, their maximum velocity would be on the order of $1 \times 10^9$ cm/s. At this speed, the ions would traverse a distance of 1 mm in ≈ 80 ps - nearly two orders of magnitude longer than the few-picosecond lifetime reported for hot, dense plasma in NWA [8]. The main difficulty with this scenario, however, is that the line emission from $Ti^{20+}$ ions would undergo a Doppler shift, $v'(1 + v\cos\theta/c) \approx v' \cdot 1.02$, where $v'$ is the photon frequency in the ion rest frame and $\theta = 45°$ is the detection angle relative to the ion velocity. For the $Ti^{20+}$ $He_\alpha(w)$ transition at 4.75 keV, this 2% Doppler shift would correspond to ≈100 eV, which was not observed experimentally.

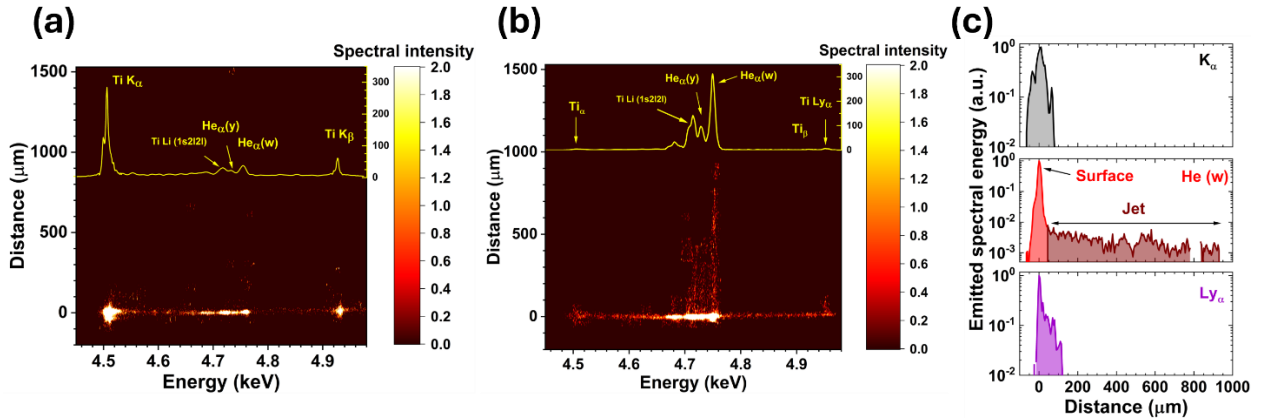

**Fig. 6.** CCD X-ray images of the Ti emission source from the reference planar target (a) and the composite NWA target (b), obtained using 1D spatial imaging. The images are intentionally oversaturated; the true emission intensity ratio between the surface and the "jet" region is ≈200. Insets show the spatially integrated X-ray spectra. (c) Spatial distributions of the spectrally integrated emission from the "cold" $K_\alpha$ line (top), the $He_\alpha$(w) line (middle), and the $Ly_\alpha$ line (bottom).

In the second scenario, we assume that $Ti^{20+}$ ions expand with the characteristic sound speed, $C_s = \sqrt{\frac{Z^* k_\beta T_e}{M_i}} \approx 3.1 \times 10^7 \sqrt{T_e[keV]Z^*/A} [cm/s^{-1}]$ where $Z^*$ is the effective ion charge, $T_e$ is the electron temperature, A is the atomic number, and $M_i$ is the ion mass. For an electron temperature of ≈ 1 keV—required to efficiently excite the $Ti^{20+}$ $He_\alpha$ (w) transition—the plasma expansion speed is $2.5 \times 10^7$ cm/s. Over 1000 μm from the target surface, this corresponds to a plasma "jet" lifetime of several nanoseconds. Thus, this second scenario suggests the formation of a keV-temperature, high-density plasma with an anomalously long (nanosecond timescale) lifetime.

## 4- Numerical simulations

To gain insight into the physics of relativistic interaction of laser pulses with NWA and flat targets, we conducted 3D Particle-in-Cell (PIC) simulations using the code QUILL [38] with Finite-Difference-Time-Domain (FDTD) Maxwell solver and Vay's pusher for electrons and ions [39]. The step size in the time and spatial domains was $\Delta t = 5 \cdot 10^{-3} \lambda/c$ and $\Delta x = \Delta y = \Delta z = 0.01\lambda$ correspondingly with $\lambda = 0.4$ μm the laser wavelength. The laser radiation is simulated by the 2 μm FWHM diameter Gaussian beam and 25 fs FWHM Gaussian pulse with the peak intensity matching the experimental value. The NWA target was simulated as an array of 25 cylinders, each with a diameter of $0.375\lambda$, spaced 400 nm center-to-center on a substrate of the size $0.75 \times 7 \times 7\lambda^3$. The reference flat target uses the same substrate dimensions as the NWA target.. The targets were assumed to be pre-ionized at the leading edge of the laser pulse to an initial electron density of $30n_{cr}$, where $n_{cr} = \frac{1.1}{\lambda^2[\mu m]} \cdot 10^{21} [cm^{-3}] \approx 6.9 \cdot 10^{21} cm^{-3}$ is the critical density, and the initial cell population is a single macro-electrons and macro-ions (macroparticles).

Snapshots of evolution of the electron density and the generated magnetic field at different selected moments of time are shown in Fig. 7. The zero is defined as the moment when the peak in the laser pulse hits the surface of the flat target or the top of nanowires for the NWA target respectively. For the flat target, the simulations predict formation of a shock wave with the density amplitude up to 120 $n_{cr}$ propagating into the target's volume (Fig. 7a). The ponderomotive electron acceleration in the laser pulse propagation direction results in the formation of the current generating the azimuthal magnetic field with the peak amplitude up to 3 kT (Fig. 7b, t = 27 fs). This direct current triggers the flow of the return current with the amplitude ≈ 0.7 kT, as follows from reversing the polarity of the magnetic field (Fig. 7b, t = 130 fs).

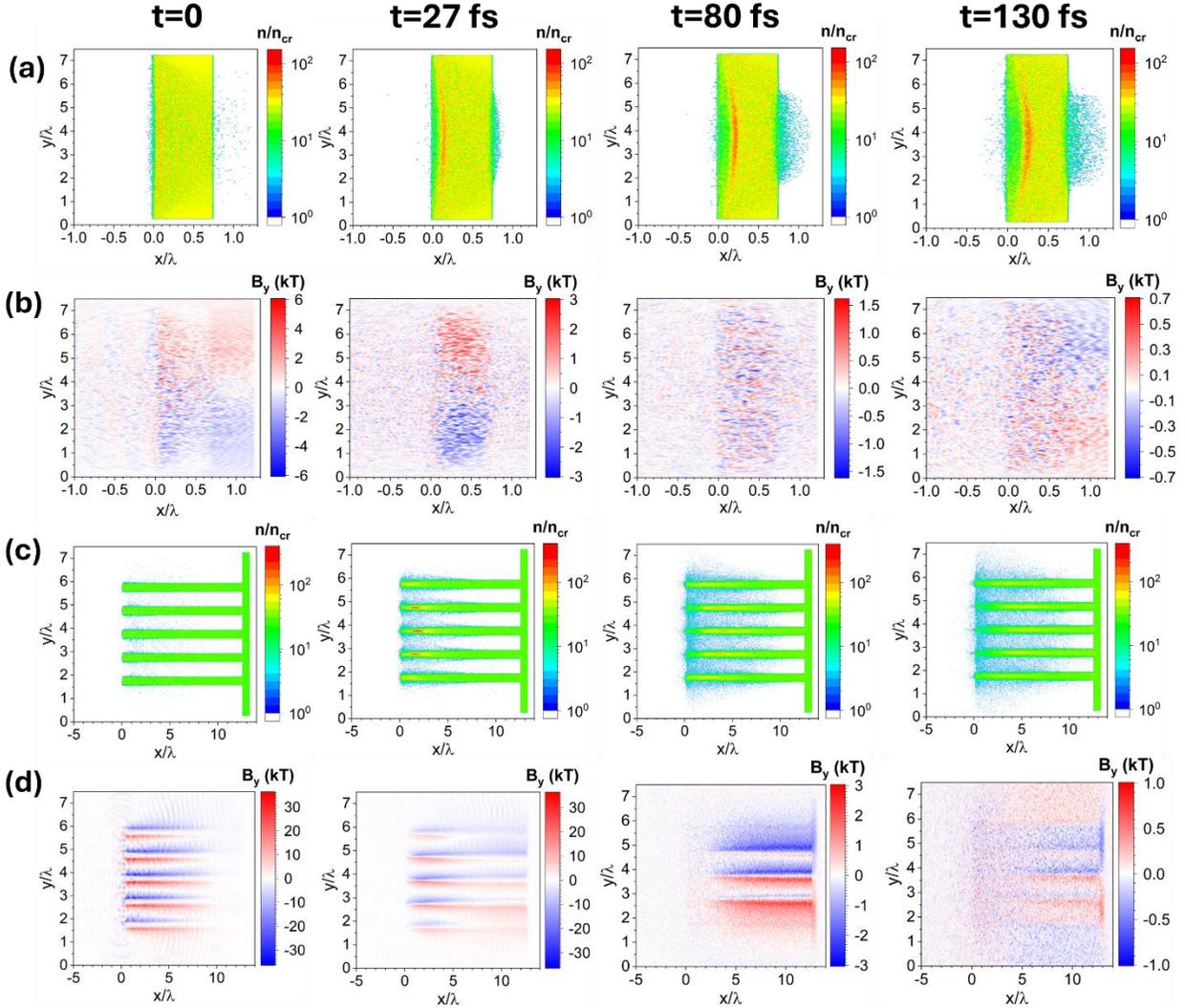

**Fig. 7** Temporal evolution of (a) electron density and (b) magnetic field in the flat reference target, and (c) electron density and (d) magnetic field in the NWA target. Electron density is normalized to the critical density, and the magnetic field is given in kT. Note the different scales of the propagation axis (x-axis) for the flat and NWA targets.

For the NWA targets, the return current sets up in each individual wire in the focal volume on a much shorter time scale than in the flat target, resulting in generation of the magnetic fields with peak amplitude up to 30 kT (Fig. 7d). This giant magnetic field leads to pinching the electron current, forming the implosion shock compressing the density up to 300 $n_{cr}$ i.e. tenfold the initial value (Fig. 7c), in agreement with predictions in [26]. Within tens of femtoseconds the interaction between the magnetic fields of individual wires results in a formation of the global magnetic field of kT amplitude that is formed by two cylindrical counter-streaming currents – the return current in the center of the focal volume and the direct current at the periphery (Fig. 7d, t=130 fs). Note that the field occupies the whole volume of the wires that is order of magnitude larger than the volume of the flat target plasma.

Complementary to the PIC simulations, we modelled the measured spatially and temporally integrated X-ray spectra using the collisional-radiative code FLYCHK [40]. The code allows implementation of two approaches. The first one is a steady-state approximation that assumes a homogeneous layer of plasma with the density and temperature remaining constant in space and time. This approach can be used even for ultrashort interaction time processes if the plasma density remains sufficiently high during the time required to reach a steady state ion charge distribution and if this time is longer than the radiative relaxation time of metastable levels [36]. The advantage of this approach is that it does not require knowledge of plasma evolution, and the plasma density and temperature can be used as fitting parameters to match the simulated and experimentally measured spectra. Thus, this approach provides an independent reference for PIC simulations. K-shell radiation emitted by highly charged ions serves as an excellent diagnostic technique for assessing the bulk electron temperature and electron density of the plasma. Analysis of emission lines from He- and H-like charge states provides the information about extreme plasma parameters reached by the end and shortly after the interaction with the laser pulse. In application to He-like $Ti^{20+}$ ions, the bulk electron temperature can be estimated from the intensity ratio between the resonance *w*-line and dielectronic satellites (DS) $1s^22l - 1s2l'2l$, whereas the plasma electron density is inferred by analyzing the intensity ratio between the intercombination *y*-line and the resonance *w*-line [36]. Following this approach, we simulated the experimentally measured spectra, shown in Fig. 5, using the plasma density and temperature as fitting parameters and convolving the synthetic spectra from FLYCHK simulations with the spectrometer resolution function modeled by a Gaussian with a FWHM of 2 eV, estimated from the spectral width of the "cold" $K_\alpha$ emission line. The results are shown in Fig. 8. Within the steady-state approximation we estimate a Ti plasma with electron density of $10^{23}\ cm^{-3}$ and bulk electron temperature of $800 \pm 50\ eV$ for reference planar targets, whereas for nanowire array the estimated electron density is one order of magnitude higher ($10^{24}\ cm^{-3}$) and a bulk electron temperature twice as high ($1600 \pm 50\ eV$). Note that higher discrepancy between the simulated and measured spectra for the planar reference target in Fig. 8a is mostly because of a low signal-to-noise ratio of the measured spectra (see Fig. 5).

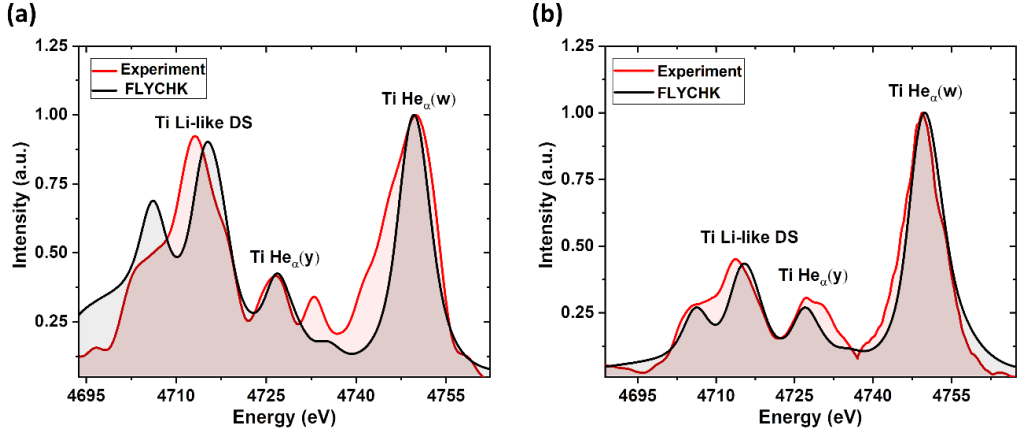

**Fig. 8.** The measured (red shaded curve) and simulated within the steady-state approximation (black shaded curve) X-ray emission spectra of He-like $Ti^{20+}$ ions from the (a) reference planar targets and (b) nanowire array targets.

The second approach is based on transient simulations where we used the history of the plasma density and temperature retrieved from the 3D PIC simulations (Fig. 9a). The time-dependent density profile was obtained by averaging the 3D density data calculated by PIC in the yz-plane (the plane orthogonal to the propagation direction) and within the 0.5 µm layer at the tip of the wires. This depth is defined by the laser energy absorption length, estimated from the PIC simulations, and by the opacity of the plasma for the line emission from He-like and H-like Ti ions. The temperature profile is obtained by fitting the Maxwellian distribution to the energy distribution function of the electrons in the low energy range (the bulk plasma temperature). The PIC simulations cover the time window of the plasma evolution up to 200 fs, therefore for the transient FLYCHK simulations we extrapolated the time dependencies of the temperature and the density calculated by PIC code up to 1 ps time scale. Also, the short time-window of the simulations suggest that the results are applicable only to the X-ray emission at the target surface, not in the "jet". An excellent match between the simulated emission spectrum of He-like $Ti^{20+}$ ions and the experimentally measured spectrum is achieved for the opacity depth of 0.2 µm and shown in Fig. 9b. Note that the plasma density estimated from the steady-state approach matches very well to the maximum value suggested by the PIC simulations, whereas the temperature is by a factor of two lower than the peak value in the transient model. The transient simulations allow also to estimate the temporal profile of the X-ray line emission shown in Fig. 9c. The dip in the temporal emission profiles is due to the plasma opacity effect. The simulations suggest that the He (w) emission lasts about 300 fs (FWHM).

Finally, our simulations allow us to make a rough estimate of the plasma parameters that can be expected in the plasma "jets" observed in the experiment. First, it is noteworthy that the spectral measurements shown in Fig. 6 are time-integrated over the whole plasma history, i.e. showing the emitted spectral energy. Therefore, we estimate the ratio of the emitted spectral energy in the jet and at the target surface as $\approx 3 \cdot 10^{-3}$ (see Fig. 6c). To find the ratio of the spectral

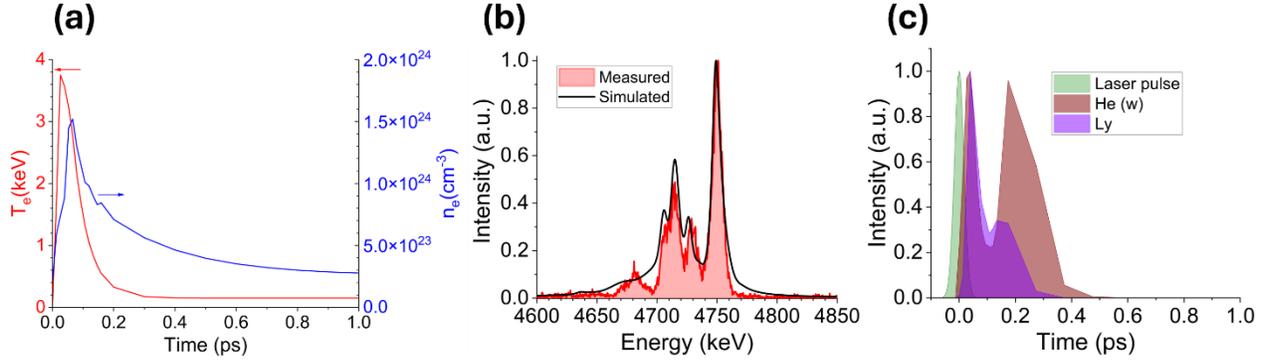

**Fig. 9.** (a) The temporal history of the plasma density (blue line, right axis) and temperature (red line, left axis) retrieved from the PIC simulations. (b) The measured (red shaded curve) and simulated (black curve) X-ray emission spectra of He-like $Ti^{20+}$ ions from the nanowire array targets. (c) Simulated time-dependent He (w) and Ly emission from He-like $Ti^{20+}$ and H-like $Ti^{21+}$ ions. The green shaded curve shows the laser pulse intensity profile.

intensities, we invoke the emission duration factor 300 fs / 3 ns for the emission source at the surface (Fig. 9c) and estimated for the jet from the ionic expansion speed. Thus, we estimate the ratio in the spectral intensity of He (w) line at the surface and in the "jet" as $\approx 10^{-7}$. Next, we run steady state FLYCHK simulations, using the electron density and the temperature as fitting parameters, to calculate the line emission spectrum in the "jet" with He (w) line intensity $10^7$ times lower than the intensity at the surface. The result is shown on Fig. 10 where the white contour marks the plasma parameters expected in the "jet". As follows from Fig. 10, to sustain the He-like emission observed in the experiments, the plasma density and temperature in the "jet" than for flat targets. This could provide long living magnetic fields of high enough strength to confine hot and dense plasma required for keeping high (He-like) ion charge state and its electronic excitation at nanosecond time scale. However, physics of Weibel instability development in the experiments of [41] remains unclear. In our case of NWA targets, simulations would require a sophisticated hybrid code that includes 3D PIC simulations for the should be in the range $5 \cdot 10^{20} - 10^{22}$ cm$^{-3}$ and 0.4 – 1.2 keV respectively. According to our PIC simulations, such values are easily accessible within the first picosecond time span of plasma evolution after interaction with the laser pulses. The intriguing question is how such parameters can be sustained at nanosecond time scale, estimated from the "jet" extend? As an explanation we propose confinement of the hot and dense plasma by a strong magnetic field generated by the plasma itself. A simple estimate based on the balance between the magnetic pressure and the gas-kinetic plasma pressure suggests that to confine the plasma with the parameters estimated for the "jet" the magnetic field strength in the range

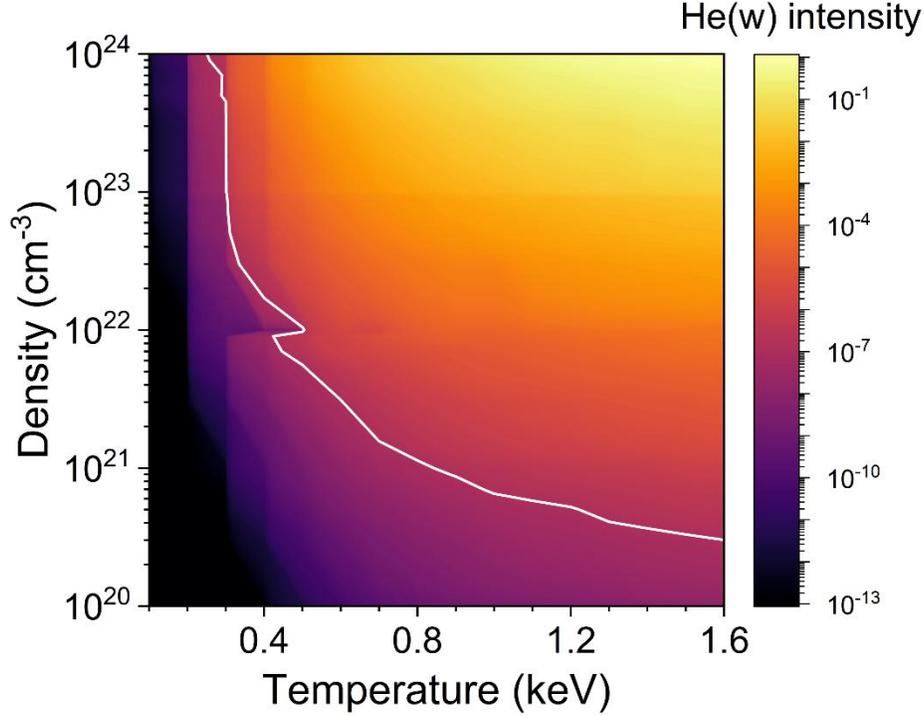

**Fig. 1.** Intensity of He-like (w) line of $Ti^{+20}$, normalized to the intensity at the target surface, as a function of plasma density (y-axis) and temperature (x-axis) simulated by FLYCHK code. The white contour marks values of plasma parameters $(T_e, n_e)$ relevant for the Ti He-like plasma "jet", i.e. when the line intensity drops by seven orders of magnitude in respect to the intensity at the surface.

200 – 600 T is required. For NWA targets, according to PIC simulations, such magnetic fields exist at least within first picosecond after the interaction with the laser pulse. However, the question of how such strong magnetic fields can be maintained on nanosecond time scale remains open. Here we note that magnetic fields with field strength ≈100 T sustained at many nanoseconds time scale were measured in recent experiments on time-resolved evolution of Weibel instability in plasmas generated by high intensity femtosecond laser pulses on flat targets [41]. Considering much larger amplitudes of the magnetic field, generated in NWA targets due to the pinch-effect in individual wires, and formation of the macroscopic global magnetic field structure on sub-picosecond time scale, we speculate that the increment and the strength of Weibel-like instability might be significantly higher in the case of laser-NWA interaction. In depth investigation of this phenomenon requires a hybrid approach involving PIC simulations for the laser pulse – NWA interaction, providing the initial conditions for a magnetic fluid code (MHD) that can handle magnetic fields with several hundred Tesla strength for multi-nanosecond timescale of plasma evolution.

## 5- Summary and Conclusions

In summary, we have presented results from detailed experimental studies of relativistic, ultrashort, ultra-high contrast laser pulse interactions with nanowire array (NWA) targets composed of Si-core/$TiO_2$-cladding structures. A comprehensive suite of diagnostics was employed, including charged-particle (electron and proton) detection and high-resolution, element-specific X-ray spectroscopy with one-dimensional imaging of the emission source. Unlike conventional approaches, our setup enabled imaging of X-ray emission extending from the target surface into the backward direction relative to the laser propagation.

We observed highly efficient backward proton acceleration via the front-surface TNSA mechanism, achieving proton energies above 5 MeV—comparable to previous reports [25]—but at laser intensities two orders of magnitude lower. X-ray spectroscopic analysis, supported by FLYCHK and 3D PIC simulations, indicates complete ionization of the nanowire volume to He- and H-like charge states, leading to the formation of volumetric plasmas with keV-scale temperatures and electron densities on the order of $10^{24}$ cm$^{-3}$. Imaging of He-like Ti$^{20+}$ emission further revealed jet-like plasma structures extending up to 1 mm from the target surface, observed exclusively with NWA targets. The scale of these structures and the high ionic charge states imply the persistence of hot (0.4–1 keV), dense ($10^{21}$–$10^{22}$ cm$^{-3}$) plasma over nanosecond timescales.

The physical mechanism enabling the confinement of such long-lived plasmas remains unresolved. We propose that strong magnetic fields, on the order of hundreds of Tesla, may provide confinement. Although both previous studies and our simulations predict even stronger fields shortly after laser–target interaction, the process by which such fields are maintained on nanosecond timescales is unknown. Experimental evidence of magnetic field amplitudes sufficient for plasma confinement was recently reported in time-resolved studies of Weibel instability in femtosecond-laser-driven flat targets [41]. We therefore suggest that a global volumetric magnetic field of kiloTesla strength, predicted to exist in NWA targets shortly after laser interaction, may seed a Weibel-like instability that generates and sustains the magnetic fields responsible for plasma confinement. The precise scenario linking these early strong fields, current-driven instabilities, and the long-term plasma evolution, however, remains an open question.

Our results demonstrate the existence of long-lived hot and dense plasmas in NWA targets, opening new opportunities for applications in laser-driven nuclear physics and the development of high-brilliance X-ray sources.

## Supplementary Material

See the supplementary material for the following: nanowire array fabrication technology; one-dimensional imaging of the X-ray emission source.

## Acknowledgements

The authors acknowledge contributions of the JETI-40 laser team (Burgard Beleites, Falk Ronneberger, and Alexander Sävert) for running the laser system and very useful discussions with Prof. Malte Kaluza and Dr. Mohammed Almassarani. The authors acknowledge the support from the BMBF project "BMBF-Projekt 05P21SJFA2" Verbundprojekt 05P2021 (ErUM-FSP T05). We also acknowledge contributions supported by COST Action CA21128- PROBONO "PROton BOron Nuclear fusion: from energy production to medical applicatiOns", supported by COST (European Cooperation in Science and Technology - www.cost.eu). The nanowire targets production was supported by German Research Foundation DFG (CRC 1375 NOA – Nonlinear Optics down to Atomic scales), project number 398816777 (Project Z3). We also acknowledge support by the German Research Foundation Projekt-Nr. 512648189.

## Author declarations:

## Conflict of Interest

The authors have no conflicts to disclose.

## Data availability:

The data that support the findings of this work are available from the corresponding author upon reasonable request.

# Supplementary materials

# Long Living Hot and Dense Plasma from Relativistic Laser-Nanowire Array Interaction

Ehsan Eftekhari-Zadeh[1,2,3,a)], Mikhail Gyrdymov[4], Parysatis Tavana[1,4,5], Robert Loetzsch[1,2], Ingo Uschmann[1,2], Thomas Siefke[6], Thomas Käsebier[6], Uwe Zeitner[6], Adriana Szeghalmi[6,7], Alexander Pukhov[8], Dmitri Serebryakov[9], Evgeni Nerush[9], Igor Kostyukov[9], Olga Rosmej[4,5], Christian Spielmann[1,2,3], Daniil Kartashov[1,2,3,b)]

[1] Institute of Optics and Quantum Electronics, Friedrich Schiller University Jena, 07743 Jena, Germany
[2] Helmholtz Institute Jena, 07743 Jena, Germany
[3] Abbe Center of Photonics, Friedrich Schiller University Jena, 07745 Jena, Germany
[4] GSI Helmholtzzentrum für Schwerionenforschung GmbH, Darmstadt, 64291, Germany
[5] Goethe-Universität Frankfurt am Main, 60438 Frankfurt am Main, Germany
[6] Institute of Applied Physics, Friedrich Schiller University Jena, 07745 Jena, Germany
[7] Fraunhofer Institute for Applied Optics and Precision Engineering IOF, 07745 Jena, Germany
[8] Institut für Theoretische Physik, Heinrich-Heine-Universität Düsseldorf, 40225 Düsseldorf, Germany
[9] Institute of Applied Physics RAS, 603950 Nizhny Novgorod, Russia

* Correspondence should be addressed to: a) e.eftekharizadeh@uni-jena.de, b) daniil.kartashov@uni-jena.de

# Nanowire arrays fabrication

The NWA targets were fabricated by electron beam lithography followed by an ICP gas chopping etching process in Silicon. The backside of the substrate is locally thinned by mask aligner lithography and subsequent ICP etching. First 30 nm of chromium is deposited by means of ion beam sputter deposition (Oxford Ionfab 300LC+) on a double side polished Silicon (Si) wafer. Afterwards a negative tone e-beam resist (OEBR-CAN038AE by TOKYO OHKA KOGYO CO., LTD.) was applied and patterned by character projection electron beam lithography (Vistec SB350OS) [1]. Next the resist pattern is transferred by ion beam etching (Oxford Ionfab 300LC+) into the chromium layer, and the resist mask is stripped by oxygen plasma etching. The actual nano wires are then generated by a gas chopping ICP etch process using $CF_4$ at -70°C (Sentech SI500). For the backside thinning the frontside of the wafer is protected by a resist layer and subsequently the backside is coated with AZ10XT (Microchemicals GmbH). This is then patterned by mask aligner lithography (EVG6300). Finally, the Si substrate is etched by ICP etching (Sentech SI500) to a residual thickness of ~55 μm. In a last step the nano wires are covered by 25 nm titanium dioxide by means of atomic layer deposition (ALD) Oxford Instruments OpAL) using Titanium Tetraisopropoxide (TTIP) and oxygen plasma precursors [2, 3].

# One-dimensional imaging of the emission source

The toroidal crystal, used in the experiments for measuring X-ray emission spectrum of Ti ions, provides X-ray diffraction and spectral focusing on the vertical plane, whereas it works as a cylindrical mirror in the horizontal plane, enabling one-dimensional imaging of the emission source. The raytracing of the corresponding 1D image for the case of isotropic, spherical plasma expansion is shown in Fig.S1. Here F is the focus of the mirror, C is the curvature center, the gray AC-line is the target surface located under 45° angle to the principal axis, A, B and C are exemplary points at the spherical plasma surface, A′, B′ and C′ are corresponding image points that should be projected at the CCD screen. As follows from the raytracing, the resulting image is a line symmetric relative to the principal axis. The raytracing for a plasma jet is shown on Fig.S2. In this case the image is in the upper plane from the principal axis.

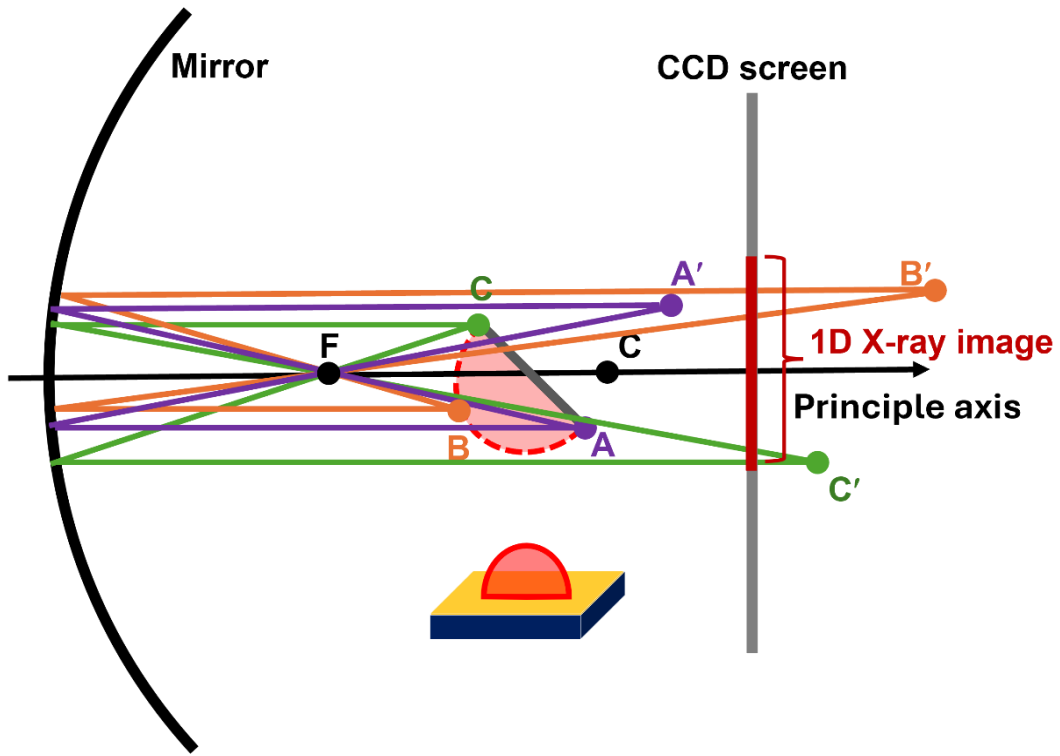

**Fig. S1.** 1D imaging for an isotropic spherical plasma expansion

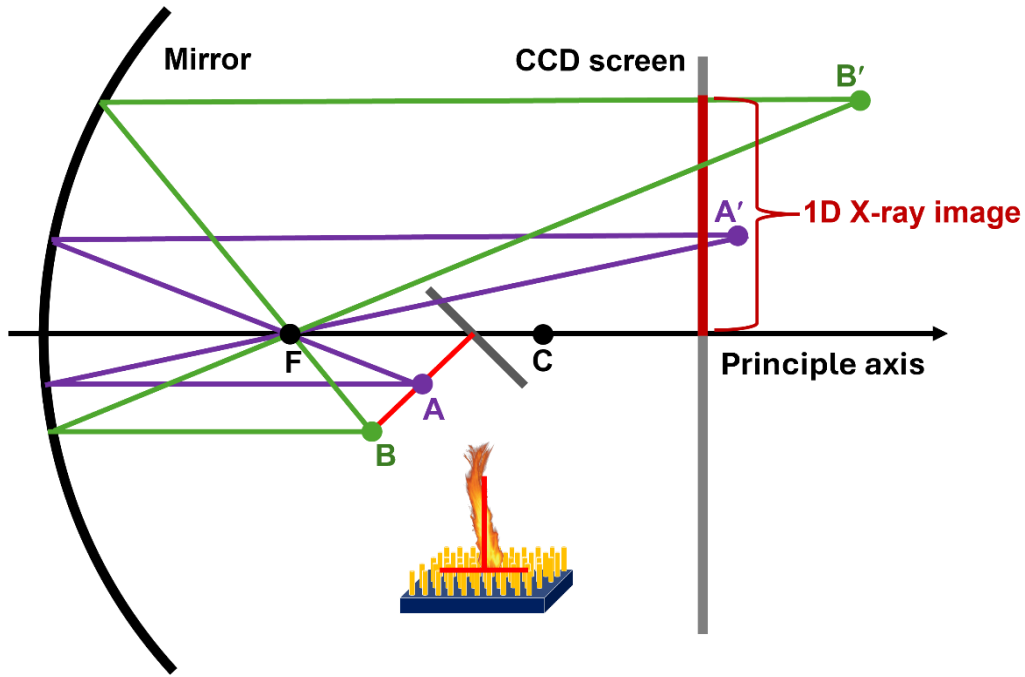

**Fig.S2.** 1D imaging for a jet

Supplementary References